\documentclass[]{spie}  

\usepackage{subcaption}
\usepackage{amsmath,amsfonts,amssymb}
\usepackage{graphicx}
\usepackage{wrapfig}
\usepackage[colorlinks=true, allcolors=blue]{hyperref}

\usepackage{xcolor}
\usepackage{ulem}
\usepackage{comment}

\title{Towards the processing, review and delivery of 80\% of the ALMA data by the Joint ALMA Observatory (JAO)}

\author{Jorge F. García Yus}
\author{Bill Dent}
\author{Drew Brisbin}
\author{Chin-Shin Chang}
\author{Laura Gomez}
\author{Theodoros Nakos}
\affil{Joint ALMA Observatory, Alonso de Córdova 3107, Santiago, Chile}

\authorinfo{Further author information: (Send correspondence to Jorge F. García Yus) J. García: E-mail: Jorge.Garcia@alma.cl, B. Dent: E-mail: Bill.Dent@alma.cl, D. Brisbin: Drew.Brisbin@alma.cl, C. Shang: E-mail: ChinShin.Chang@alma.cl, L. Gomez: E-mail: Laura.Gomez@alma.cl,  Th. Nakos:  Theodoros.Nakos@alma.cl}

\begin{document} 
\maketitle

\begin{abstract}
After eight observing Cycles, the Atacama Large Millimeter-submillimeter Array (ALMA) is capable of observing in eight different bands (covering a frequency range from 84 to 950 GHz), with 66 antennas and two correlators. For the current Cycle (7), ALMA offers up to 4300 hours for the 12-m array, and 3000 hours on both the 7-m of the Atacama Compact Array (ACA) and TP  Array plus 750 hours in a supplemental call. From the customer perspective (i.e., the astronomical community), ALMA is an integrated product service provider, i.e. it observes in service mode, processes and delivers the data obtained. 

The Data Management Group (DMG) is in charge of the processing, reviewing, and delivery of the ALMA data and consists of approximately 60 experts in data reduction, from the ALMA Regional Centers (ARCs)  and the Joint ALMA Observatory (JAO), distributed in fourteen countries. Prior to their delivery, the ALMA data products go through a thorough quality assurance (QA) process, so that the astronomers can work on their science without the need of significant additional calibration re-processing. 

Currently, around 90\% of the acquired data is processed with the ALMA pipeline (the so called pipeline-able data), while the remaining 10\% is processed completely manually. The Level-1 Key Performance Indicator set by the Observatory to DMG is that 90\% of the pipeline-able data sets (i.e. some 80\% of the data sets observed during an observing cycle) must be processed, reviewed and delivered within 30 days of data acquisition.

This paper describes the methodology followed by the JAO in order to process near 80\% of the total data observed during Cycle 7, a giant leap with respect to approximately 30\% in Cycle 4 (October 2016 - September 2017). 

\end{abstract}


\section{INTRODUCTION}
\label{sec:intro}  
\subsection{ALMA in a nutshell}
\label{subsec:nutshell}
ALMA is a large international observatory that carries out astronomical observations at millimeter and sub-millimeter wavelengths. ALMA's specified frequency range is 35 to 950 GHz\footnote{The full frequency range covered by ALMA will be achieved once Bands 1 and 2 have been successfully commissioned.}, corresponding to an approximate wavelength range of 8.5 mm to 0.3 mm. Because these frequencies are strongly absorbed by clouds and water vapor in the atmosphere, the instrument is located in the very dry Atacama region of northern Chile. The location of the observatory is the Chajnantor plateau, which is at an altitude of 5000 meters above sea level. 

The array consists of parabolic antennas that work together to simulate a single telescope. Despite the antennas being quite similar in appearance compared to those used for lower-frequency radio astronomy and for communications, they are specially designed to be extremely stiff and to have a surface accuracy better than 12 microns while operating in an extremely hostile environment. There are:
\begin{itemize}
    \item Fifty dishes of 12-meter diameter to form an extended array, which form the 12-m array, whose signal is combined through the baseline (BL) correlator,
    \item 12 dishes of 7-meter diameter to form a close-packed array (the Atacama Compact Array, or ACA, also known as the Morita array\footnote{Dedicated in remembrance of Professor K.-I. Morita}), whose signal is combined through the ACA correlator, and 
    \item Four dishes of 12-meter diameter, that are used primarily as single-dish telescopes.
\end{itemize}
 The antennas in the 12-meter array can be moved around on the plateau to provide different configurations, ranging in extent from about 150 meters up to 16 kilometers.

\subsection{Description of the ALMA observing cycles}
On every October 1st, ALMA initiates a new observing cycle. The current cycle is Cycle 7 and it was expected to last until the end of September 2020. However, due to the COVID-19 contingency, a decision was taken to extend Cycle 7 until the end of September 2021. Thus, Cycle 8 will officially start on October 1st, 2021. 

A new cycle does not only mark the start of the acquisition of data coming from new proposals, but also the use of a new ``on-line'' software i.e. a new version of all software components involved in data acquisition, new observing capabilities not previously offered to the astronomical community (such as longer baselines, higher frequency observations, etc) and a new pipeline, capable of processing both legacy and possibly (even if partially) new observing modes, offered to the community with the new observing cycle.   

Most of the ALMA data are processed through the ALMA pipeline, released as part of CASA (Common Astronomy Software Applications)\footnote{For more information on CASA, see \url{http://casa.nrao.edu}}. Specific observing modes, such as solar, VLBI, and full polarization, produce data that have to be processed manually, rather than through the ALMA pipeline. Independently of how a dataset has been processed, though, it will go through a Quality Assurance process, to ensure that the data products that will be delivered to the Principal Investigators (PIs) meet the expected quality standards, in terms of sensitivity and angular resolution (AR), are free of observational and instrumental issues, and their corresponding calibration follows the observatory guidelines. 

In the case that the generated products comply with the PI's sensitivity and AR, they will be ingested to the ALMA Archive, setting a proprietary time of 12 or 6 months, depending on the type of the proposal. After that, the data become public to the astronomical community, through the ALMA science archive.

\section{The Data Management Group}
The Data Management Group is in charge of the processing of all ALMA data. Although organically DMG belongs to the Department of Science Operations (DSO) at the JAO, in its wider sense DMG comprises data processing experts at the JAO, the ALMA Regional Centers (ARCs) and the ARC-nodes.

The structure of the ARCs and ARC-nodes is as follows:
\begin{itemize}
    \item EA ARC : consists of the East Asian ALMA Regional Center (EA ARC) at the NAOJ Headquaters, Mitaka, Japan, and nodes at ASIAA, Taipei (Taiwan), and at KASI, Daejeon (Republic of Korea).
    \item EU ARC : consists of the European ALMA Regional Center (EU ARC) at the ESO Headquarters, Garching, Germany, and its system of Nodes at  Ondrejov (Czech Republic), Grenoble (France), Bonn/Cologne (Germany), Bologna (Italy), Leiden (Netherlands), Onsala (Sweden), Manchester (UK), and an expertise centre in Lisbon (Portugal).  
    \item NA ARC : consists of the North American ALMA Regional Center (Charlottesville, VA, USA) and the Canadian ARC-node in Vancouver (BC). 
\end{itemize}

In total, there are approximately 60 data reducers (DRs) around the world who process and review ALMA data.

\subsection{Level-1 Key Performance Indicator for the Data Management Group}
The Level$-$1 Key Performance Indicator (KPI) set by the Observatory to DMG is the following: 90\% of the pipeline-able data sets should be delivered within 30 days of when they become Fully Observed (FO). This time-goal is relaxed to 45 days for the data sets that can only be processed manually. 

The pipeline-able data correspond to the so-called standard observing modes, that are sufficiently well characterized, both in terms of data acquisition and data processing, so that the observations can be successfully calibrated and imaged using the ALMA data reduction pipeline. The data which do not go through the pipeline must be manually calibrated and imaged, and both tasks are extremely expensive in terms of person-time. 

Considering that (1) at least 90\% of the data observed throughout a full Cycle go through the pipeline and that (2) ALMA observes some 3000 Member Obs Unit Set (MOUS) per Cycle, it is therefore expected that DMG processes (through the ALMA pipeline) and reviews some 2400 MOUS per Cycle, and that each of these data sets  are delivered within 30 days of data acquisition. 

It is estimated that some 2400 data sets will be processed at the Santiago data center, located on the JAO premises, where a limited number of  9 Data Reducers (DRs) review and deliver this crucial majority-number of data sets. The ARCs are in charge of the labor intensive duty of manually processing the data that do not go through the pipeline and contributing to the processing and reviewing of the pipeline-able data sets not taken by JAO.

\subsection{The DMG as part of an ecosystem}
\label{sec:ecosytem}

\begin{figure}[t]
 \centering
 \includegraphics[width=66mm]{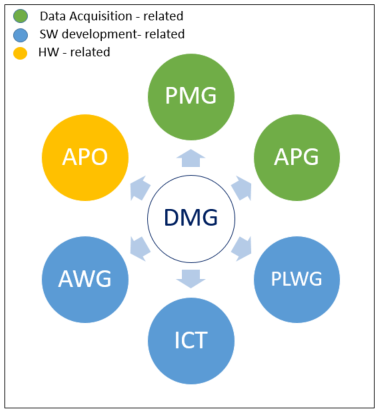}
 \caption{The DMG ecosystem: The Data Management Group interacts with the Program Management Group, the Array Performance Group, the Pipeline Working Group, the Integrated Computing Team, the Archive Working Group and the Archive and Pipeline Operations Group. The different colours indicate the main activity the group is focused on: green refers to activities related to data acquisition, blue for software development and orange when the main focus is on hardware.}
 \label{ecosystem}
\end{figure}

Because of the nature of DMG's work, our staff is involved in many more activities other than data processing and QA. The DMG ``ecosystem'' \cite{nakos}, i.e. the groups we are interacting with, are listed below:
\begin{itemize}
    \item Program Management Group (PMG), in charge of the data acquisition. The relationship between PMG and DMG is bidirectional: PMG informs DMG about possible issues that might affect data processing and DMG gives feedback about the quality of the processed data, so as to avoid the execution of poor-quality observations.
    \item Array Performance Group (APG). APG is in charge of the acquisition of data corresponding to new observing modes, using the data acquisition software (also known as ``on-line'' software) that will be used for the forthcoming Cycle. DMG has to process these datasets, evaluate their quality, and report back to APG. 
    \item Pipeline Working Group (PLWG). DMG provides requirements to the PLWG regarding improvements, new features, reporting problems or bugs and provides general feedback and context on necessary pipeline refinements. Furthermore, approximately a month before the beginning of a new Observing Cycle, the PLWG delivers to DMG the version of the ALMA pipeline that will be installed in production at the beginning of the new Cycle and will be used to process the new-cycle ALMA data. DMG must run the pipeline both at JAO and the ARCs, and compare the results with a benchmark of approximately $10-15$ datasets (provided by the PLWG) to ensure that all sites obtain the same results. Once the validation has finished successfully, JAO officially accepts the new version of the pipeline, which is deployed in production and is used to process data from the new observing Cycle.   
    \item Integrated Computing Team (ICT). Updates of the ALMA software components not directly related to data acquisition go to production every two months (with some exceptions for specific months). Furthermore, there are on-going projects related to software development, which have a direct impact on data processing. DMG interacts with the ICT members (mainly, but not exclusively, with the developers and the release managers, both at JAO and the ARCs) on a near-daily basis, to create tickets for new features, coordinate the forthcoming activities (software deployment, project advancement). Additionally, DMG performs integration testing of the features that will be included in the future software releases, in coordination with the ICT members. 
    \item ALMA Science Archive (ASA). All ALMA data which pass the QA2 criteria are ingested in the ASA. DMG is the main internal (within ALMA) stakeholder in the Archive Working Group, which is in charge of the maintainability, curation and evolution of the ASA.
    \item Archive and Pipeline Operations (APO) group, at JAO. APO is in charge of the processing infrastructure (processing nodes and disk space), the hardware infrastructure related to ASA and the associated data bases, among other tasks. DMG interacts on a daily basis with APO colleagues, for discussing operational matters related to data processing, ingestion of the ALMA data in the archive, etc.
\end{itemize}
A schematic representation of DMG's relationship with the above groups is presented in Fig.~\ref{ecosystem}. More details about some of the components of the DMG ecosystem can be found in ~\cite{espada14}.

\section{Structure of the ALMA data} 

The organization of a science project is subdivided into several hierarchical levels. At the bottom of this structure are the Scheduling Blocks (SBs), the minimum set of instructions describing an ALMA observation. An SB contains a large amount of information about what should be observed, including, but not limited to, the position and velocities of the science targets, frequency settings, frequency coverage  and  resolution,  angular  resolution and details of the correlator setup.

Each SB typically consists of (a) set-ups, (b) calibrations, and (c) target observations that can be observed within 1.5 hours. The end of an SB execution may be specified either (a) in terms of the maximum allowed time on the sky or (b) when certain well-deﬁned science goals have been reached. An SB cannot be stopped and re-started. Therefore, an SB either runs to completion, fails, or is terminated by the Astronomer on Duty (AoD)\footnote{An Astronomer On Duty, or simply AoD, is an astronomer in ALMA's control room who monitors the executions of the different SBs and evaluates, in real time, the quality of the data obtained}. Given the limited duration of the SBs, it is often necessary to observe them several times to achieve the sensitivity and angular resolution required by the Principal Investigators (PIs). 

Each repetition of an SB generates a new Execution Block (EB); the sum of the EBs required to reach the PI goals are grouped into a Member Observing Unit Set (MOUS). DMG's deliverables to the astronomical community are properly calibrated and imaged MOUSs. 

More information regarding the data structure of the ALMA data can be found in the ALMA technical handbook~\cite{remijan2019}.

\section{Quality assurance (QA)}
\label{sec:qa}
The objective of the ALMA quality assurance (QA) is to ensure that the data products meet the PI-specified goals within  cycle-specific  tolerances. That is, the delivered products have reached the desired control parameters outlined in the science goals (or are as close to them as possible), they are calibrated to the desired accuracy, and calibration and imaging artifacts are mitigated as much as possible.  

To be more efficient in detecting problems, ALMA QA has been divided into several stages that mimic the main steps of the data flow. The broad classification\index{Quality Assurance!classification} of this multi-layered QA approach is: 

\begin{description}
\item[QA0]
The first stage of QA, known as QA0, deals with performance parameters on timescales of an SB execution length or shorter, and is thus performed at the time or immediately after each execution, i.e. per EB. The initial QA0 assessment is performed by the AoDs at the OSF, using the software tools AQUA, aoscheck, and QuickLook (more on these tools in section~\ref{tpi}). This information is complemented with reports derived using the Monitor \& Control (MC) display tools, which monitor specific parameters not directly tracked by the calibrations (e.g., total power level variations, weather parameters, etc). QA0 metrics/parameters have been selected to check the health of the whole signal path from the atmosphere down to the correlators, as well as issues connected with the observation itself. Combined with additional checks in AQUA itself, QA0 performs a thorough  quality assessment for each EB. More detailed information on some of these tools can be found in Chavan et al ~\cite{chavan16}.
\item[QA1]
In addition to data taken during the SB execution, the observatory tracks several performance parameters of both the array as a whole and the individual array elements (i.e. antennas). These parameters, which vary slowly (typically on timescales longer than a week) and can potentially affect the data quality, are measured by AoDs and the System Astronomers at predefined periods as ``observatory tasks''. Additionally, they are monitored after major interventions, including, but not limited to, antenna moves, front-end swaps, sub-reflector repairs, or if significant deterioration of performance is detected during operations. Standard procedure requires that, after the completion of an intervention of such an invasive character, the specific array element will not be used for regular science observations. Only when the antenna integration has been completed successfully, and specific array and antenna parameters have been measured and have been implemented in the system, can the antenna be included in the array for taking PI data.  
\item[QA2]
Quality assurance of the data products generated once the observations have been processed falls under the purview of QA2, deals with the quality assurance of the data products generated once the observations have been processed with either the ALMA data reduction pipeline or through the so-called script generator, which is used for manual processing. The ALMA pipeline and the script generator are built on CASA, which is the official software for ALMA data processing and analysis. It should be noted that the ARCs do not provide support to users for any data processing or analysis done outside of CASA\footnote{JAO does not directly interact with the astronomical community, thus all helpdesk tickets are channelled through the ALMA Regional Centers.}. The ALMA pipeline and script generator are run both at JAO and the ARCs. It is only during this stage of data reduction that the science goals set by the PI can be compared with the actual values in the data products (i.e., RMS, angular resolution, signal-to-noise ratio (SNR) and dynamic range).
\item[QA3]
Post-delivery evaluation of the data products sent to the PIs is part of QA3. It is advisable that PIs check the data products themselves very soon after delivery. The QA3 process is either triggered by a PI or any astronomer who has downloaded public data, who has detected some issue in their data which may reflect an underlying problem with the observing procedure, the data themselves, the processing or the ingestion of the data products in the ALMA archive, or by the observatory itself, once it has identified a problem affecting a specific or multiple data sets. It should be noted that the data products of observations which are under QA3 cannot be downloaded from the archive (although the raw data can be obtained, upon request).   
\end{description}

\section{From the sky to the PI}
\label{tools}
This section provides additional information about the tools and processes involved in the data-acquisition, data transfer, data processing and data ingestion phases. Given the importance of QA0 and QA0+ in the current work, a subsection is dedicated to these tools.

\subsection{Data-acquisition related software tools}
\label{tpi}

The following software components are involved in the data acquisition process, either throughout the duration of an observation or after its end, when an EB is generated: 
\begin{itemize}
    \item Telcal: this is the so-called ``online"\footnote{In the ALMA terminology the term ``online" refers to any process related to data acquisition, while the term ``offline" refers to the processes related to data processing.} telescope calibration software, which evaluates in real time the health of the entire system, performs the calibrations, forwards information to other software components and stores the results in the ALMA archive.
    \item Quick-look: this tool allows to visualize several parameters related to the observations such as the precipitable water vapor (PWV), the phase variations, the antennas included in the array, etc.
    \item aoscheck: this software component assesses the quality of each newly-generated EB and decides on its status (QA0\_PASS, QA0\_SEMIPASS or QA0\_FAIL). It yields information for outliers and generates warnings on potential hardware or software issues. aoscheck estimates key values for making the assessment for the EB, based on the number of operational antennas, the spatial resolution, and the percentage of the executions completed.
    \item QA0+ script: Based on aoscheck's input, the QA0+ script does a basic reduction of the continuum data, computes statistics from the calibrations and creates images of the targets. These results are then displayed in the Alma QUality Assurance tool (AQUA) and   saved to the archive tables. In the case aoscheck can not make a recommendation on the status of a specific EB, the AoD will use the QA0+ information to make the assessment.  
\end{itemize}

\subsection{QA0 and QA0+ in steps}
The aim of QA0 is to rapidly identify problems during or soon after each observation which will affect the final data quality. Problems can include issues with the observing system (such as antenna or correlator faults), issues with the calibration targets, with the data integrity, or with weather conditions (such as unstable atmospheric phase). Rapidly identifying such issues within minutes of an observation being completed enables the array operator to, for example, fix or remove antennas, change to a different calibration target, or to move to a different observing band. ALMA is a multi-element interferometer: it can continue operations with some antennas removed for repair, or change to a different frequency to better suit the atmospheric conditions. The main goal is to identify the problem fast.

Addition of new QA0 checks and flags can be done with a relatively fast turnaround time - typically 1-2 months (or faster, if critical). So if a new hardware problem is recognised, it is possible to include additional data checks relatively rapidly (this has happened several times in the last few years).

QA0 only provides checks of the calibration data - it does not look at the data on the science target nor do imaging. The amount of science data is typically 3-4 orders of magnitude more than that of the calibration, and trying to analyse it would make the QA0 too slow and cumbersome. Identification of problems in the science data, a second phase of QA0 (known as QA0+) provides a stripped-down imaging pipeline using CASA with python. QA0+ does some very basic calibration (for WVR correction and phase calibration), applies the QA0 flags, and transforms the visibilities into an a-priori-calibrated image of the continuum on the science target. No spectral data is available, for reasons of speed and disk usage. The QA0+ pipeline typically takes one hour. The image is checked for distortion due to adverse weather conditions, and the target flux and image noise are compared with expected values. If multiple observations of the target have been done previously, it will combine them to obtain both the rms and synthesized beam. Although it takes considerably longer than QA0, it also provides the PI with feedback by the next day. This is often necessary for checks of fluxes for Target of Opportunity observations, such as supernovae, GRB and comets.

\subsection{From the OSF to the Santiago Central Office, to the ARCs}
A temporary archive stores all raw data at the OSF, for six months. However, once an observation is completed, its binary data will be transferred and stored in the Next Generation Archive System (NGAS) at the Santiago Central Offices (SCO). Similarly, the metadata will be transferred and stored in the Oracle database at SCO. These transfers take place using a dedicated fiber link and a radio link for backup.  The flow is replicated from SCO to the three ARCs, safely saving up to four copies of each observation within a few hours (unless there are issues with the transfer associated with the internet service  providers).

Once at SCO, the data are processed in the main data center: over 45 high performing nodes (256 GB of RAM, with 8$-$12 cores per node) and a high performance Lustre{\textregistered} storage system provide access to CPUS, memory and disks with a very low latency (thanks to the INFINIBAND switches). 

In case an MOUS passes all quality control criteria (see subsection~\ref{sec:qa}) it must first be ingested in the NGAS at the SCO, independently on where the processing or review took place. Once ingested, it will be replicated at the ARCs by a software component called ``Product Ingestor'' a dedicated service that guarantees that the data can be accessed from all the ARCs.

\subsection{The MOUS life-cycle}
\label{lifecycle}

\begin{figure}[htbp]
\centering
\includegraphics[width=.5\linewidth]{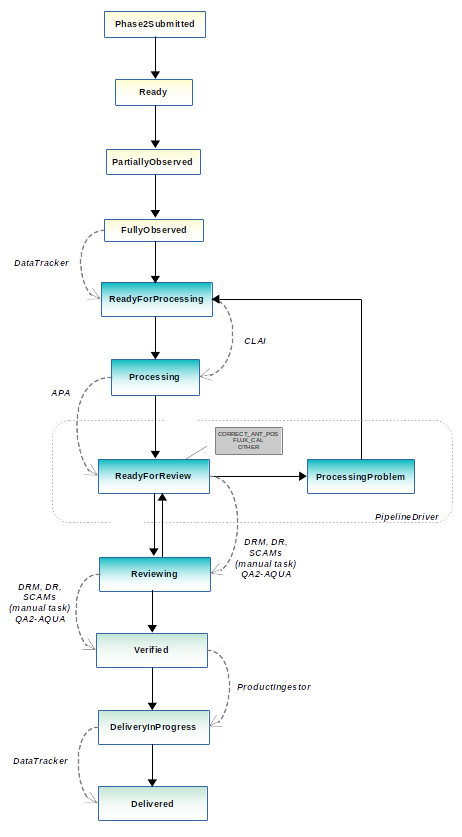}
\caption{Presentation of the so-called life-cycle of a Member Observing Unit Set (MOUS).\\
a) ObsUnitSet transits from Phase2Submitted to Ready\\
b) ObsUnitSet becomes PartiallyObserved\\
c) Member ObsUnitSet becomes ReadyForProcessing when its SchedBlock becomes FullyObserved\\
d) ObsUnitSet becomes Processing\\
e) ObsUnitSet becomes ReadyForReview when Pipeline/Manual script has finished successfully\\
f) ObsUnitSet becomes Reviewing when a Pipeline/Manual Reviewer checks the outcome of the product\\
g) ObsUnitSet becomes Verified when the reviewer finds that imaging is successful\\
h) ObsUnitSet becomes DeliveryInProgress when the data package has been ingested into the JAO Archive, and replication to the ARCs\\
i) ObsUnitSet becomes Delivered when the corresponding data package has been delivered to the PI's ARC, and the PI notification email has been sent\\}
\label{workflow}
\end{figure}

The ALMA data follow a strict path from the moment an SB is prepared by the PI (see figure~\ref{workflow} for more details), starting with the scheduling sub-system that decides which SB should be executed, based on the array compatibility, weather conditions, project ranking, visibility of the targets or any other time constraints, etc. Each time that an EB is finished, many services, like instruments in an orchestra, will work in a coordinated way to produce results. 

When the sufficient amount of data is collected for a given SB, the MOUS shall be considered as ``FullyObserved'' and will automatically transition to a state called ``ReadyForProcessing''. Subsequently, the Cluster ADAPT Interface (CLAI) Tool, which runs at the JAO data center \footnote{Actualy CLAI runs both at JAO and the ARCs.}, will queue the dataset for processing, using the appropriate routines among a list of predefined processing recipes (for interferometric or single dish data). A successfully processed MOUS transitions to the ``ReadyForReview'' state, and is therefore available for going through the QA2 process by a highly-trained DMG personnel at JAO or at the ARCs or ARC-nodes, through a browsable interface.

Once a DR becomes available, she/he will assign the MOUS to herself/himself for review through the AQUA interface. The MOUS state will automatically change to “Reviewing”. The data product quality will be assessed according to the QA2 criteria corresponding to the specific cycle. In case the DR considers it necessary, the MOUS can be sent back to the processing queue, to include some manual flagging\footnote{``Flag'' is astronomy jargon for data cleaning, artifact removal, outlier removal, etc.} to correct issues that the pipeline heuristics could not detect. The options for the QA2 assessment are QA2\_PASS, QA2\_SEMIPASS or QA2\_FAIL.

\section{Operational Analysis}

\subsection{Multiple entries in the MOUS life-cycle and their implications}
Following up on what was presented in subsection~\ref{lifecycle}, setting an MOUS state to QA2\_SEMIPASS or QA2\_FAIL is a necessary but costly action for the observatory, and the cost will depend on what part of the life-cycle it happens. More specifically:
\begin{itemize}
    \item Even if an MOUS is sent back to the observing queue observations might not take place if the observing conditions are not optimal.
    \item Re-observing an SB consumes precious observing time that could be used for observing other projects.
    \item Returning an SB back to the observing queue is not always possible, in case the requested array configuration is no longer available.
    \item  Reprocessing increments the nodes occupancy and of course translates into additional person-time for reviewing the same data set, diverting resources that could be used to process, review and deliver other data sets. It also translates into possible delays in meeting the 30-days KPI. 
\end{itemize}

\subsection{Issues diagnosis}
\begin{figure}[htbp]
\centering
\includegraphics[width=134mm]{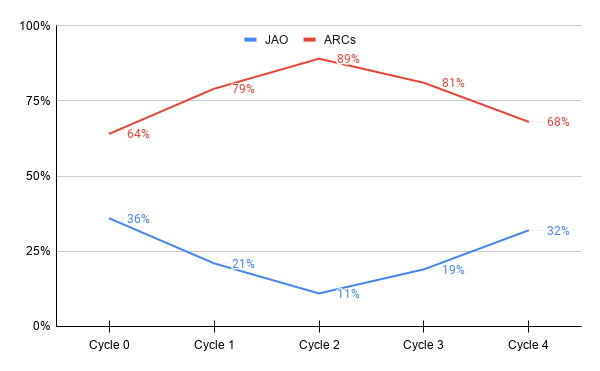}
\caption{Time evolution of the data processing load for JAO (blue) and the sum of the three ARCs (EA, EU and NA, in red) as a function of observing Cycle. The two percentages should always give 100\%. Starting with Cycle 2 as a reference, a 10\% increment in the percentage of datasets processed at JAO translates into JAO meeting the goal to process 80\% of the datasets produced by a given Cycle by Cycle 9.}
\label{percentage}
\end{figure}

The time evolution of the percentage of data sets processed at JAO and the ARCs, between Cycles 0 and 4, is presented in Figure~\ref{percentage}. It should be clarified that between Cycles 0 and 2 all data were processed manually (i.e. both the calibration and image data products were done manually), and it is only in Cycle 3 that the pipeline (calibration only) was deployed in production. This pipeline calibrated the majority of the ALMA datasets, but the imaging still had to be performed manually. In Cycle 4, though, a fully automated pipeline (calibration and imaging) became available. 

Based on Figure ~\ref{percentage}, slightly more than 30\% of the ALMA data obtained during a whole Cycle was processed at JAO by Cycle 4. According to the trend (increment of 10\% per year), it would take about five additional ALMA cycles (equivalent to almost five years) before JAO processes 80\% of the ALMA data (i.e. by late 2022, as Cycle 4 ended in September 2017).

\begin{table}[t]
\begin{center}       
\begin{tabular}{|c|c|c|c|}
\hline
\rule[-1ex]{0pt}{3.5ex} \bf Cycle & \bf BL(hours) & \bf ACA(hours) & \bf Antennas offered (BL/ACA/TP) \\
\hline
\rule[-1ex]{0pt}{3.5ex}  2 & 2,000 & 2,000 & 34/9/2 \\
\hline
\rule[-1ex]{0pt}{3.5ex}  3 & 2,100 & 2,100 & 36/10/2  \\
\hline
\rule[-1ex]{0pt}{3.5ex}  4 & 3,150 & 1,800 & 40/10/3  \\
\hline
\rule[-1ex]{0pt}{3.5ex}  5 & 4,000 & 3,000 & 43/10/3 \\
\hline
\rule[-1ex]{0pt}{3.5ex}  6 & 4,000 & 3,000 & 43/10/3 \\
\hline
\rule[-1ex]{0pt}{3.5ex}  7 & 4,300 & 3,750 & 43/10/3 \\
\hline
\end{tabular}
\end{center}
\caption{Summary of observing hours (Baseline and ACA correlators) and number of array elements offered per observing Cycle.}
\label{table:timeoffered}
\end{table}

A more careful analysis actually revealed that this prediction was too optimistic, as it did not consider the increment in the offered hours on the sky for the following cycles. Based on the information provided in Table~\ref{table:timeoffered}, there will be some 37\% increase in the offered hours for the BL array in Cycle 7, compared to Cycle 4 (see subsection~\ref{subsec:nutshell} for the array definitions). For the ACA, the increase is more than 100\%. Roughly speaking, the increase in the expected number of data sets to process scales linearly with the increase in the number of offered observing hours. 

Furthermore, one should also consider the increase in the number of 12-m array elements that will be included in the BL array observations, between Cycle 4 and the ``steady state'' operations phase, when the minimum number of antennas to be included will be much higher (see column 4 in Table~\ref{table:timeoffered}). \\
An increase in the number of antennas, though, translates into an increase in the size of the data  proportional to the number of baselines, i.e. $N$ x $(N-1) / 2$ (where N is the number of array elements). Based on the number of 12-m antennas offered for the BL array in Cycle 4 (40) and Cycle 7 (43), there will be a 16\% increase in the number of baselines. The increase in the size of the data directly impacts the processing time, which means that data might potentially start piling up in the processing queue, as there will no processing nodes available.

Considering the Cycle 4 situation and all elements presented above, the DMG team at JAO took a series of actions in order to both increase and consolidate its processing capacity. 

To reach such a high level of processing performance, it became clear that the most relevant factors were the following:
\begin{itemize}
    \item The maximization of the number of MOUSs that finish their processing without errors,
    \item The minimization of the number of MOUSs that go through the pipeline multiple times, and 
    \item The minimization of the number of MOUSs that needed any type of manual intervention after the end of the pipeline processing (manual flagging, additional imaging after the pipeline processing, etc).
\end{itemize}

\subsection{Collecting information}  

\subsubsection{Data mining}  

The official platform used in ALMA operations for the problem report, issues and projects tracking is JIRA\MakeUppercase{\textregistered}. For the purpose of the analysis performed to optimize DMG's performance in general, and more specifically that of JAO, three projects  were deemed relevant, as they provided useful information about issues that arose between data acquisition and delivery. More specifically, these projects were:

\begin{description}
\item[PRTSPR (Problem Report and Tracking System):] 
Although initially this project was only used for reporting problems that occurred during observations, DMG promoted its use for the reporting of problems affecting the data flow downstream (i.e after data acquisition). These tickets have high visibility and can be assigned to any group, including external contributors, through a reallocation or cloning to another JIRA\MakeUppercase{\textregistered} environment.
\item[APO (Archive and Pipeline Operation):] 
The issues reported in this project are related to problems, improvements and bugs related to the processing cluster infrastructure, problems related to specific processing nodes or the overall system performance, pipeline and software administration, issues with data ingestion(s) or problems with the ALMA databases.
\item[SACM (Science Archive Content Manager):] 
The official name for the DRs at JAO is Science Archive Content Managers. The SACM tickets are mostly used for (a) reporting issues related to data quality, (b) recording actions such as the return of an MOUS back to the observation queue, (c) reporting the processing or reversing the quality assurance assessment of a MOUS. In general, these tickets are associated to very expensive decisions for the observatory in terms of observing time, processing resources and person-power.   
\end{description}
This analysis referring to more than 17,000 tickets collected and analyzed, with the support of the open source machine learning package Orange 3, used for clustering the data and visualization. Aggregating the information in categories was by no means a trivial task, considering that:
\begin{itemize}
    \item The information came from a non-structured language,  
    \item There were plenty of typos, syntax or grammar errors,
    \item The classification could not be easily done per cycle, as the use of a mandatory prefix in the subject, was only present since Cycle 4 on-wards, and, finally,
    \item The issues reported had to be classified based on the stage of the QA process at which the ticket was reported, i.e. QA0, QA1, QA2 or QA3.
\end{itemize}

\subsubsection{Issues classification}
\label{subsubsec:ic}
The beginning of Cycle 5 marked an important milestone in DMG's operational model, with the fully automated implementation in production of the concept of the MOUS life-cycle (the details about this concept have been provided in subsection~\ref{lifecycle}). The life-cycle implementation was assisted through the interaction of several complex software and hardware components. The analysis allowed issues to be grouped in categories based on areas of interest:

\begin{description}
    \item [AQUA:] The main tool for the QA assessment of all ALMA data (both at EB and MOUS level).
    \item [Archive:] Issues relevant to metadata storage and data replication at the ARCs.
    \item [Pipeline:] Bugs or any other pipeline-related issues (at software level) that prevent the pipeline from successfully terminating all processing stages for a given pipeline recipe.
    \item [Procinfra:] Issues related to the processing infrastructure, such as processing cluster malfunction, performance degradation, Lustre{\textregistered} (or file system) issues, issues related to the resource management (torque), processing node allocation or task-scheduling availability.
    \item [DARED (currently CLAI):]
    DARED was the software component allocating the processing recipes (interferometric, single-dish, 12-m array, 7-m array) according to the type of data to be processed. By June 2020 (Cycle 7) DARED was replaced by a software component called CLAI.     
    \item [NGAS:]
    Hardware infrastructure for saving the bulk binary data in real time, and mirroring at the Santiago data center and the ARCs.
    \item [Product Ingestor:]
    This is the software component in charge of the ingestion of the products into the ALMA Science Archive. Product Ingestor issues are related to problems ingesting and replicating data products from the JAO to the archives at the ARCs.
\end{description}

\subsubsection{Prioritization}
\label{subsubsec:prior}
\begin{figure}[t]
\centering
\includegraphics[width=134mm]{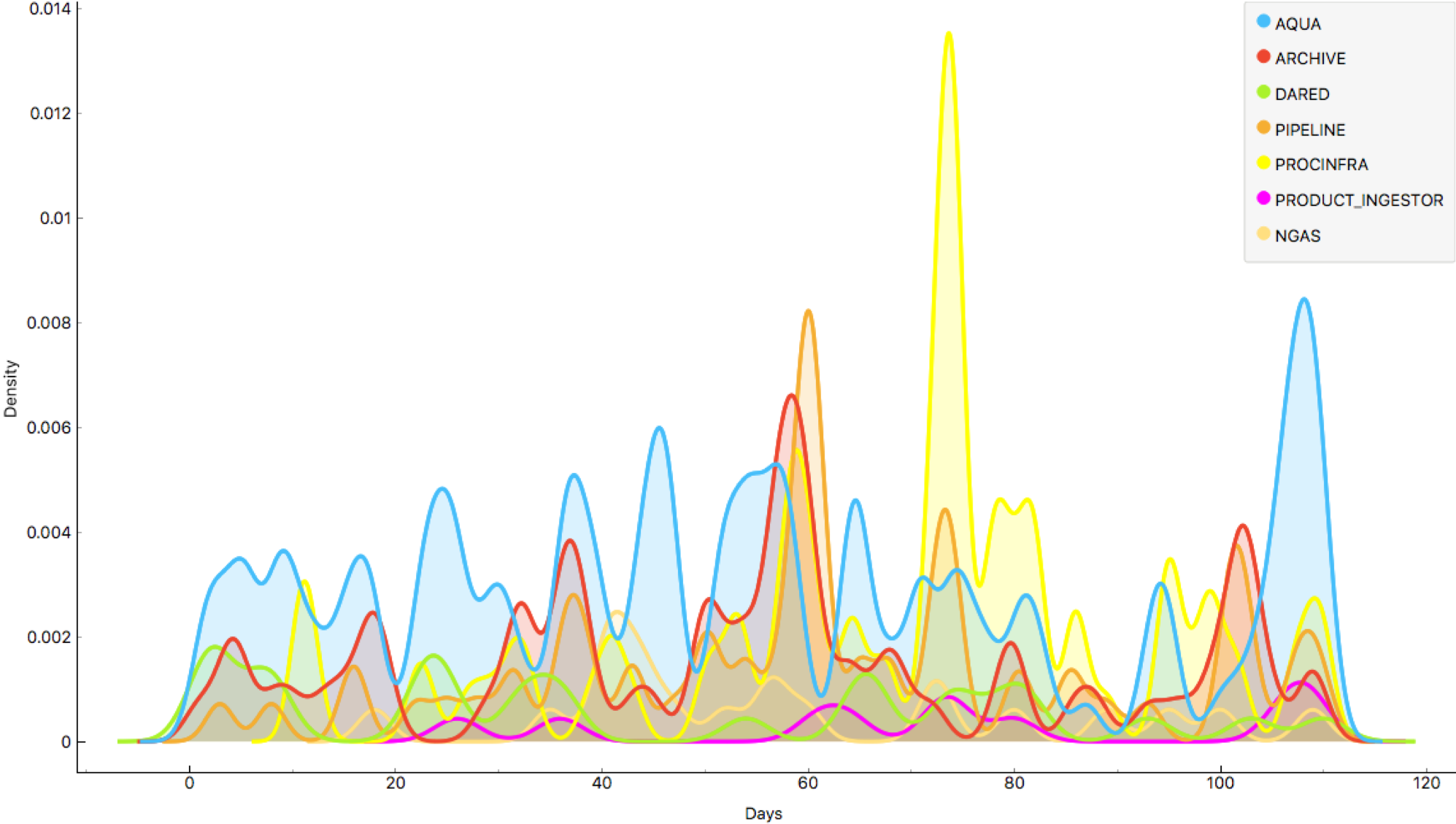}
\caption{Time evolution since the beginning of Cycle 5 (October 2017) until end of January 2018 (i.e. four months), of the different categories of issues affecting DMG's performance.}
\label{Jira}
\end{figure}

\begin{figure}[htbp]
\centering
\includegraphics[width=144mm]{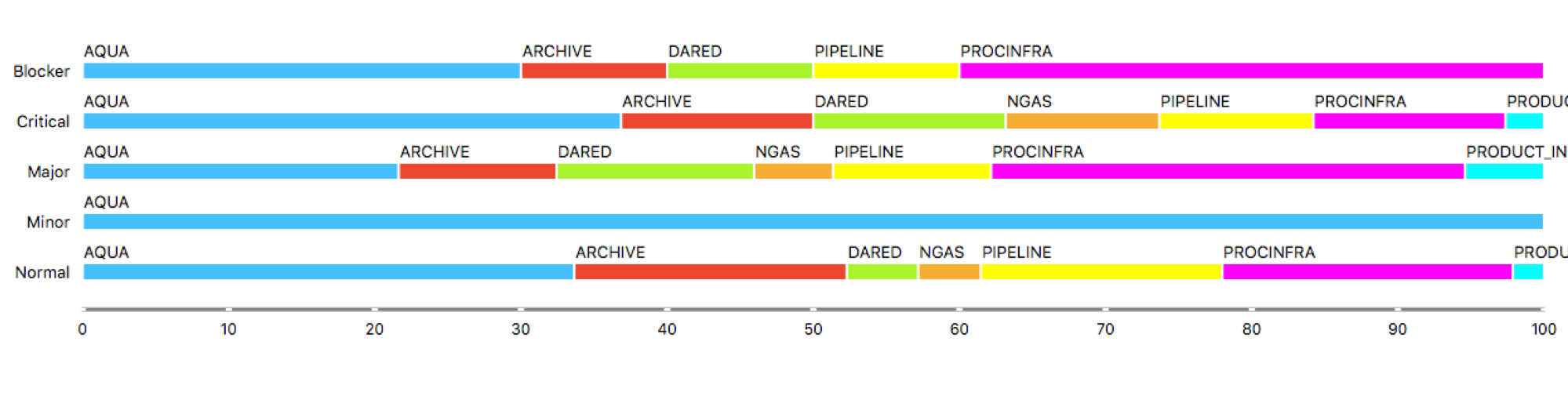}
\caption{Classification of the different issues affecting DMG's performance classified according to the ticket severity (minor, normal, major, critical, blocker).}
\label{severity}
\end{figure}

The beginning of Cycle 5 (October 2017) was marked by a high number of reported incidents related to several issues identified during the introduction of the fully automated life-cycle concept. The time evolution between October 2017 -- end of January 2018 (i.e. four months), of the different categories of issues presented in the previous subsection (\ref{subsubsec:ic}), is shown in Figure~\ref{Jira}. The vertical axis is given in units of ``ticket density'', i.e. number of tickets per day, per category, normalized by the total number of tickets over four months, of the specific category. The distribution of these categories per severity grade (i.e. normal, minor, major, critical, blocker) is presented in Figure~\ref{severity}).

From the graph it became evident that although specific components, such as the processing infrastructure (``Procinfra'' did not have a high constant presence in terms of tickets, they did show some spikes in specific dates (e.g. between dates 70 - 80 since October 1st, 2017, for the case of ``Procinfra''), implying that there were some components which were fragile and needed specific corrective actions. On the other hand, specific software components, such as AQUA and Archive, showed both spikes and a high number of tickets throughout the full four month period.

\section{Mitigation}
\label{sec:mitig}

Systematically reporting all instances where issues were found provided the opportunity to:
\begin{itemize}
    \item Raise visibility of the most common recurring problems and estimate their impact.
    \item Take data driven decisions, i.e. allocate resources for troubleshooting based on the impact of the issues (severity, number of data sets affected), and
    \item Learn from our mistakes.
\end{itemize}

In the following sections the different actions that took place during Cycles 5 and 6 are presented in detail. 

\subsection{Cycle 5 focus: tool consolidation}
\label{subsec:c5}

Based on the previous analysis, there were a series of actions taken since the beginning of Cycle 5 (October 2017) which were successfully completed by October 2018 (beginning of Cycle 6):

\begin{description}
\item [Problem reporting:] The JIRA\MakeUppercase{\textregistered} reporting system was enriched with new features. New categories, labels and tags were added for describing problems, inviting the reporter to provide more specific information. The ID of the EBs or MOUSs affected can now be linked to the relevant tickets, allowing reporters to both estimate the impact of an issue and facilitate the troubleshooting. Another mandatory information requirement introduced was the identification of the stage in the operational chain where the problem was discovered. This helps restrict the possible root causes of the issue. Finally, all tickets were automatically sent to a pre-assigned domain specialist, reducing the response time when an issue occurred. 
\item [Infrastructure monitoring:]
The processing nodes, the Lustre{\textregistered} file system, and the disk space availability are now permanently monitored by industry standardized monitoring systems such as Nagios and Grafana. Automated alerts are triggered to notify of possible future failures (so as to take preventive actions), and when a failure occurs, so that corrective actions are taken at the earliest possible time. 
\item [Infrastructure maintenance:]

Emphasis was given to establishing a robust hardware and software setup which facilitates replacement of the computing nodes. Measures include redundant hardware such as power supplies and double networks (active/passive) on one centralized operating system image server.\\
On the Lustre{\textregistered} file system, managed farm disks allow adding new storage nodes, thus ensuring hardware redundancy and active/passive metadata servers.\\
For the HPC hardware; nodes, cpus and core numbers are adopting one special configuration aligned with the pipeline requirements.
\item [Pipeline:]
Significant effort was put into improving the communication with the Pipeline Working Group (see section ~\ref{sec:ecosytem}), including the introduction of new requirements and feedback regarding the pipeline performance. Input from stakeholders helped to improve the procedures at different levels. Expertise from JAO on error debugging also ameliorated several technical issues related to the pipeline process. These efforts help mitigate the cost of reprocessing data and optimize the use of resources (both hardware and person-power). 
\item [Upstream quality assessment - issue identification and characterization: ]
Several measures were implemented to promptly identify data quality issues after an EB is completed. Within one minute of EB completion, the results from a set of calibration statistics are available for analysis. Problems with individual antennas, receivers, mixers, or scans are identified at two levels: minor (in which case the array operator is notified) and major (in which case data flagging is required, at the data processing stage). The final QA0 state is set based on this assessment, ensuring the generation of optimal quality data (both at the telescope level and in terms of data ingested into the archive).
\item[Upstream quality assessment - knowledge transfer:]
A finely tuned operational chain (data acquisition - data processing - data ingestion) requires that the two initial links of the chain, the array operators and the AoDs, have a deep knowledge of the system, so that, when necessary, they make the best decisions from an operational point of view. Renewed effort has been given to improving the available documentation and procedures, and providing in-depth training in aspects such as data structure, the software components involved in data acquisition and hardware. As a result, the number of EBs that would not be evaluated by aoscheck and QA0, and which needed an additional assessment by the data reducers at JAO, dropped from some 20\% of the observed EBs to less than 5\%.
\item [Upstream quality assessment- automation:]
Significant effort has been put into maximizing the number of issues that were automatically captured, the obvious benefit being the availability of more resources for data processing and QA2 assessment. 
\end{description}

By the end of Cycle 5 the statistics demonstrated that JAO had processed some 50\% of the ALMA data collected over the full cycle, i.e. much better than the expected 40\% according to the trend in Figure~\ref{percentage}. Furthermore, 90\% of the pipeline-able data sets that were delivered throughout Cycle 5 (some 1667 MOUS), were delivered within 50 days, a huge improvement compared to the Cycle 4 numbers (a factor of 2, approximately). 

\begin{figure}[t]
\centering
\includegraphics[width=134mm]{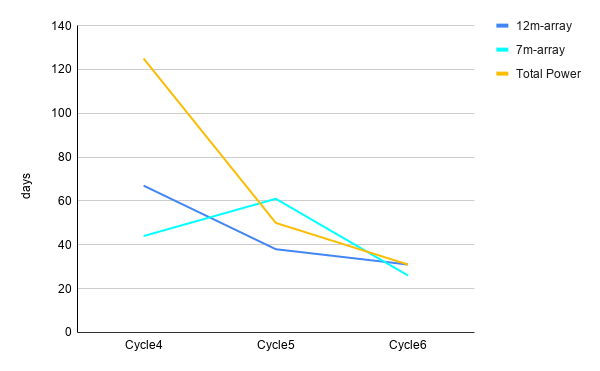}
\caption{Number of days between data acquisition and delivery of 90\% the data collected during Cycles 4, 5 and 6, separated per array type.}
\label{fig:deltime}
\end{figure}

Figure~\ref{fig:deltime} presents the improvement in the time required to deliver 90\% of the data throughout Cycles 4, 5 and 6, separated by array type (see  section~\ref{subsec:nutshell} for the array definitions). From the graph it becomes evident that for all three arrays the delivery times have improved significantly in Cycle 6 (Oct. 2018 - Sept. 2019) compared to Cycle 4 (Oct. 2016 - Sept. 2017).  

Thanks to the improved methodology, and the coordinated actions taken with the support of stakeholders from the DMG ecosystem, by the end of Cycle 5 the data processing ``machinery'' became much more stable. There were very few failures related to the processing infrastructure and the issues reported about ingestion or other software components had dropped by approximately 80-90\%.

\subsection{Cycle 6 focus: process optimization}

Throughout Cycle 6 the weight of the analysis shifted from the \textit{consolidation} of the performance of the DMG hardware and software tools  towards the \textit{optimization} of the DMG processes. More specifically, the objective was to map the QA0$-$QA2 domain at a finer resolution (i.e. at EB and MOUS level, respectively) and identify the reasons that caused EBs/MOUSs to have their QA0 status reverted from QA0\_PASS to QA0\_SEMIPASS or QA0\_FAIL, or were set to QA2\_FAIL. As previously mentioned, classifying data as SEMIPASS or FAIL translate into (a) loss of observatory time, (b) significant person-power spent on the quality assessment and (c) unnecessary use of computing resources (which, depending on the array configuration, might indeed be scarce). 

\begin{figure}[t]
\centering
\includegraphics[width=134mm]{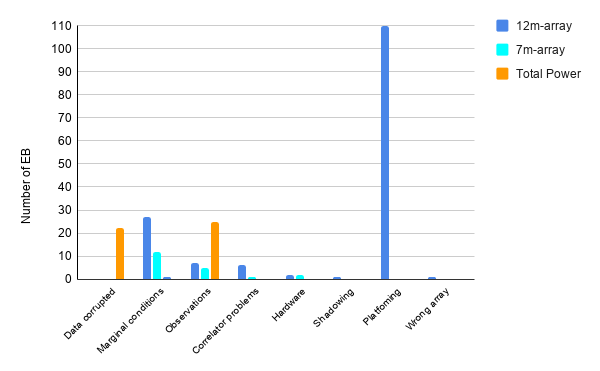}
\caption{Number of EBs versus reason for which Execution Blocks were reversed from QA0\_PASS to QA0\_SEMIPASS, \textit{after the QA2 assessment was performed}. The reasons are separated by array type.  For the EBs counted here processing resources were spent calibrating the data and producing images, and additionally a data reducer had to perform quality assessment on the data, only to find out that there was nothing that could be done to save those data. During Cycle 6 DMG took specific actions to automatically identify all those problems upstream, i.e. at QA0 level, rather than at QA2.}
\label{fig:qa0}
\end{figure}

\begin{figure}[t]
\centering
\includegraphics[width=134mm]{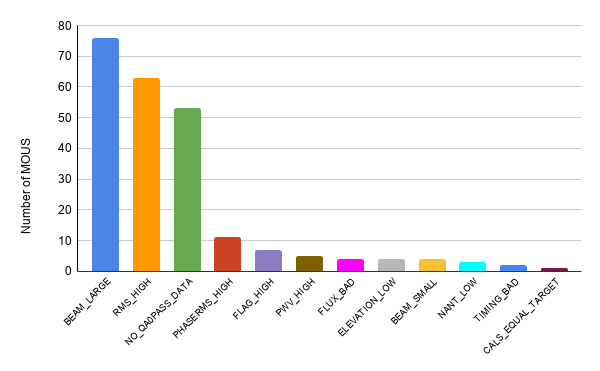}
\caption{Number of MOUSs versus reasons for which the specific MOUSs were sent to QA2\_FAIL, during Cycle 6. From the histogram it becomes clear that the first three reasons (beam large, high RMS and no QA0\_PASS data) dominate the distribution of the failures. Consequently, DMG took specific actions to mitigate the impact of these reasons on the MOUSs collected over an ALMA cycle.}
\label{fig:qa2}
\end{figure}

Figures~\ref{fig:qa0} and \ref{fig:qa2} graphically show the type of plots used for identifying the causes of the failures at QA0 and QA2 level, respectively. Without going into many details, both plots present the reasons for which data did not comply with the expected quality standards. For both cases the vertical axis corresponds to number of data sets (EBs for  fig.~\ref{fig:qa0} and MOUSs for fig~\ref{fig:qa2}). Based on the number of data sets affected, priorities were set and actions were taken to mitigate these issues.

\section{Results}

As the observatory shutdown its operations in March 2020 due to the COVID-19 pandemic, between 60$-$65\%  of the observations planned during the current Cycle (C7) have been accomplished (1934 MOUSs, with a typical number over a Cycle of the order of 3000$-$3100). 
The results presented in this section can therefore be considered as representative, with the caveat that projects that need long baseline array configurations have not been observed. Although the data produced when observing at long baselines can be processed with the ALMA pipeline, it is possible that they might need additional manual imaging, thus they are considered more laborious in terms of person-power. These projects are observed during the austral winter season, when the atmospheric conditions are the driest and the phase stability is exceptionally good. 

\begin{figure}[htbp]
\centering
\includegraphics[width=134mm]{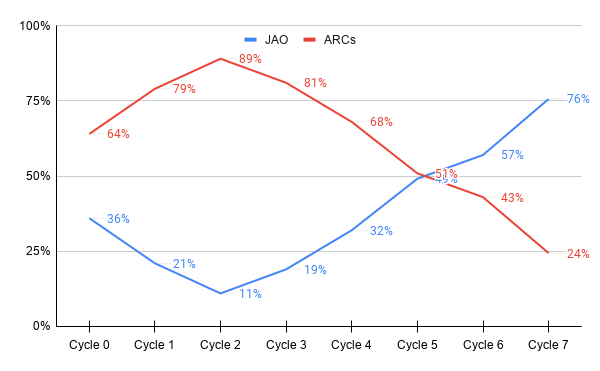}
\caption{Load processing evolution per site (as in Fig. \ref{percentage}).}
\label{fig:procload}
\end{figure}

\begin{figure}[t]
\begin{subfigure}{0.5\textwidth}
\includegraphics[width=0.8\linewidth, height=4cm]{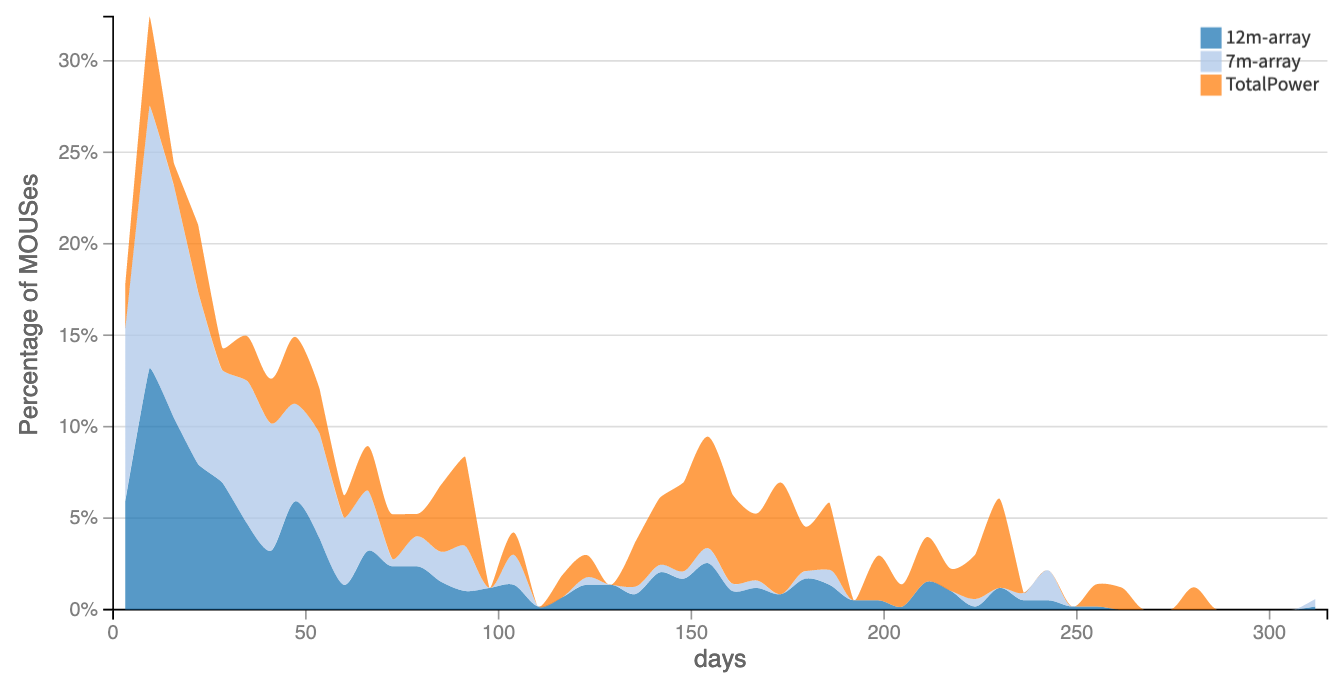} 
\caption{Cycle 4, 67/44/125 days}
\label{cycle4}
\end{subfigure}
\begin{subfigure}{0.5\textwidth}
\includegraphics[width=0.8\linewidth, height=4cm]{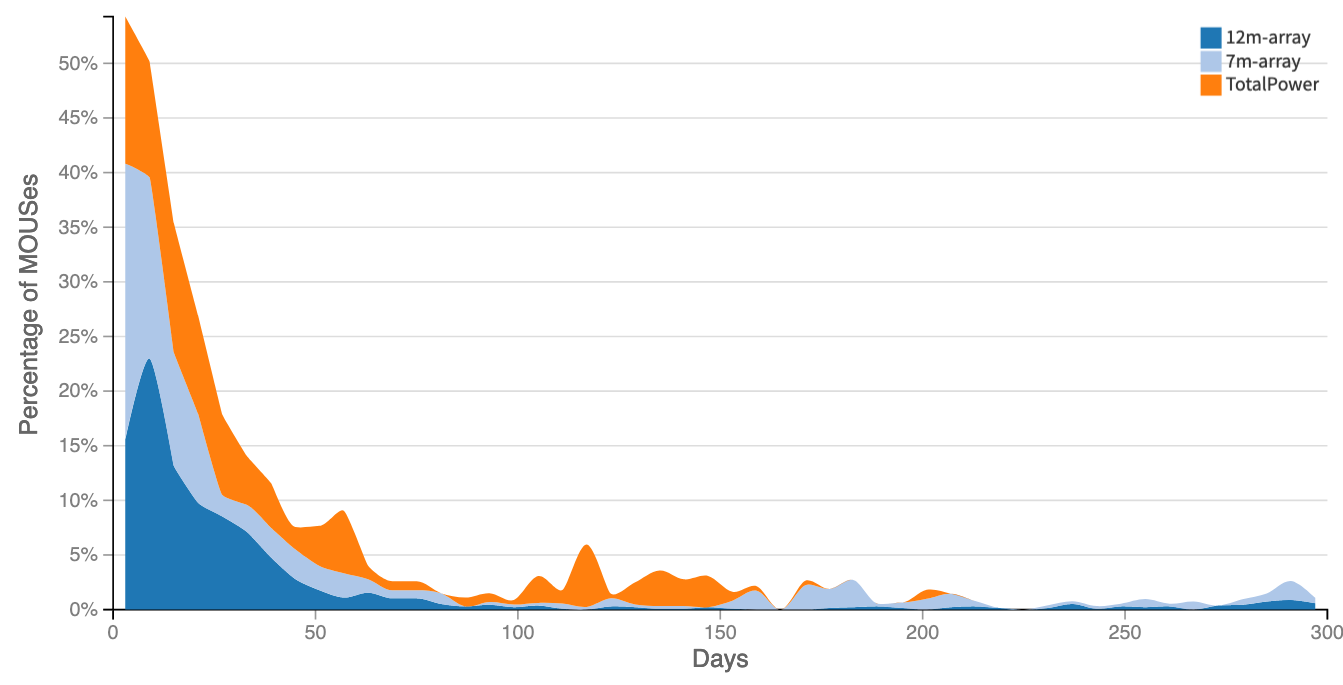}
\caption{Cycle 5, 38/61/50 days}
\label{cyecl5}
\end{subfigure}
\begin{subfigure}{0.5\textwidth}
\includegraphics[width=0.8\linewidth, height=4cm]{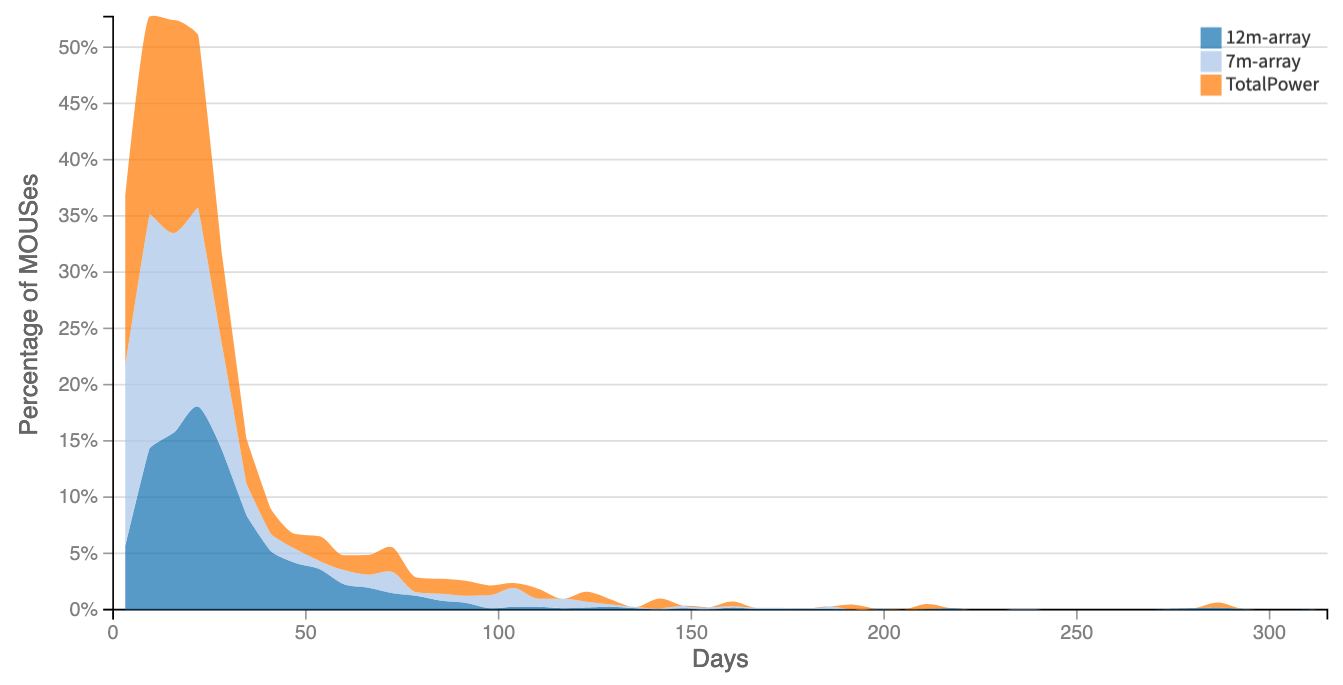}
\caption{Cycle 6, 31/26/31 days}
\label{cyecl6}
\end{subfigure}
\begin{subfigure}{0.5\textwidth}
\includegraphics[width=0.8\linewidth, height=4cm]{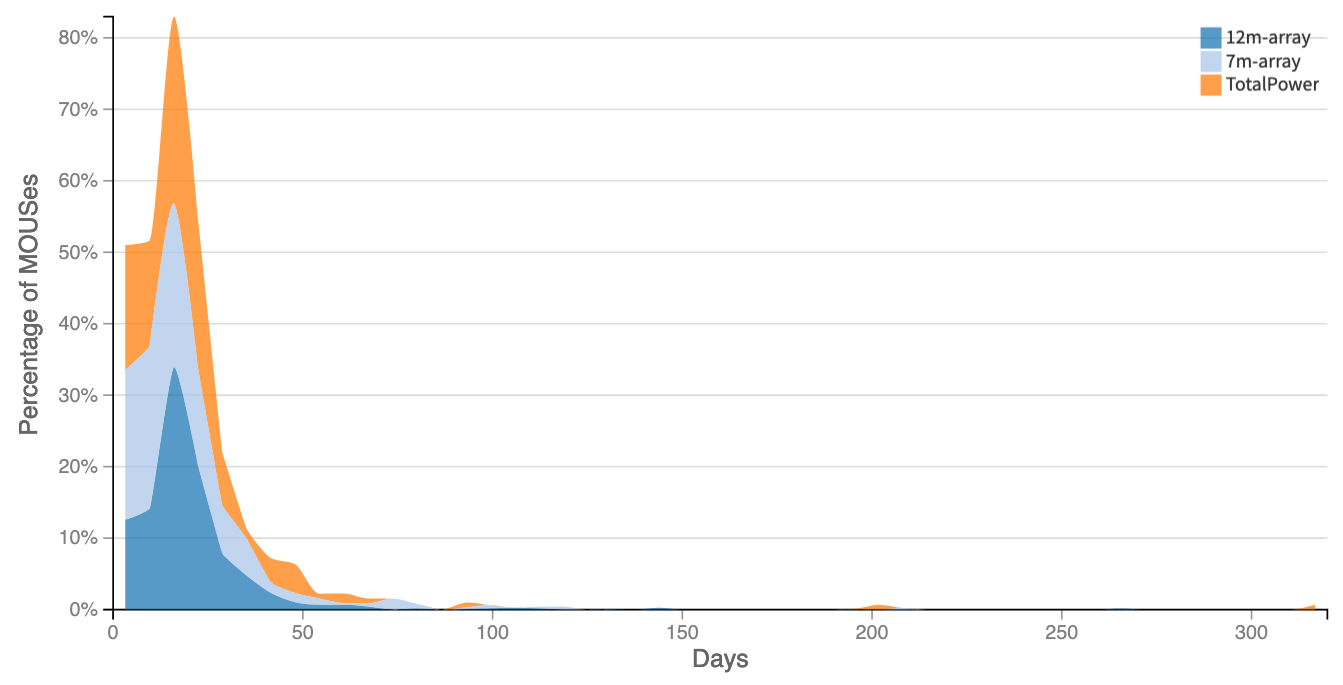}
\caption{Cycle 7, 20/18/20 days}
\label{cyecl7}
\end{subfigure}
\caption{Mean data delivery time, separated by array type (BL/ACA/TP), between Cycles 4$-$7. The actions taken have allowed DMG to decrease its delivery times for the BL, ACA and TP array, from 67, 44 and 125 days in Cycle 4, to just 20, 18 and 20 days in Cycle 7. }
\label{c4567}
\end{figure}

Figure~\ref{fig:procload} is the extension of figure~\ref{percentage}, when including Cycles 5, 6 and 7. The plot demonstrates that DMG has been processing up to 76\% of the completed Cycle 7 observations, a small percentage away from achieving the future goal of processing 80\% of the ALMA data obtained throughout an observing Cycle. Figure~~\ref{c4567} presents the average delivery times, separated by array type, for Cycles 4, 5, 6 and 7. This graph leaves no doubt that the actions taken have dramatically improved DMG's performance locally and globally (i.e. both at JAO and the ARCs).  

\section{Conclusions}

This paper presents the data-driven methodology and the subsequent actions taken to move JAO towards our goal of processing 80\% of the data in just a few Cycles, a dramatic improvement over just 32\%  being processed by JAO at the end of September 2017. The approach to face this challenge was based on the concept that the DMG deliverables to the astronomical community are cultivated as entities in a complex ecosystem, where all the processes and teams need to interact in a coordinated and synergistic manner, helping us reach our key performance indicators. If any of the subsystems operate without this coordination, their behaviour will have an impact on the other subsystems, and therefore on us.  

There are three major lesson learnt from this work. 

Quality assurance criteria can not be written in stone for such complex and continuously evolving systems. This is especially relevant for ALMA, which offers new observing modes with new observing cycles. Therefore, the quality criteria must be continuously evaluated, and at the same time the areas for improvement  (individual processes or more complex work/data flows) should also be reviewed regularly. Here we have discussed the experience of reversing the QA0 status of execution blocks only at a very late stage of the process. It has also happened that data that were QA2\_PASS and were delivered to the PIs came back as individual or massive QA3 cases. Depending on the stage where the issue is found (QA0, QA2 or QA3) and its magnitude (one, a few or many EBs / MOUSs) there is a negligible/small/huge impact on the observing, processing, or person-power budget, but regardless of the stage there is an undeniable impact.

Having a well structured file reporting system is essential. Practicing reverse engineering is expensive although the existence of natural language processing (NLP) is making this task easier. There is no doubt, however, that it will always be more efficient and precise to extract information from a database where all possible issues are described in a consistent way, the issues are associated with the affected observations and they contain well defined keywords and tags, used to characterize common problems.

One aspect that has not been discussed in this article, but is pertinent to note, is the demand for processing nodes. During regular operations there are two instances that can cause serious impact on getting the tasks done: 
a) The demand for nodes for testing purposes, which in the DMG's case, and during specific periods can affect up to 25\% of the resources, and 
b) The appearance of a massive QA3 request.
These kind of demands outline a new scenario, where the operational resources dedicated to the current cycle data processing and the resources needed for testing or reprocessing are not sharing the same infrastructure. It is worth considering the possibility to make use of cloud services, as they offer the possibility of adaptable resources, based on the daily demand.  

Finally, a deep knowledge of all the subsystems and their interactions are essential and so is the definition and the common understanding of all relevant metrics. DMG's KPIs are well defined and our performance is regularly monitored to check whether we are on or off-course.

\appendix    

\acknowledgments 
 
The authors of the paper would like to express their enormous gratitude to all past and current DMG members for their dedication and commitment to DMG's mission. Similarly, DMG would like to express its heartfelt thanks to all teams which form part of our Ecosystem and the individuals we are interacting with. Without their constant effort and professionalism we would have never reached this level of performance.   

ALMA is a partnership of ESO (representing its member states), NSF (USA) and NINS (Japan), together with NRC (Canada), MOST and ASIAA (Taiwan), and KASI (Republic of Korea), in cooperation with the Republic of Chile. The Joint ALMA Observatory is operated by ESO, AUI/NRAO and NAOJ.

\bibliography{report} 
\bibliographystyle{spiebib} 
\end{document}